\documentclass{article}

\usepackage[final]{nips_2018}

\usepackage[utf8]{inputenc} % allow utf-8 input
\usepackage[T1]{fontenc}    % use 8-bit T1 fonts
\usepackage{hyperref}       % hyperlinks
\usepackage{url}            % simple URL typesetting
\usepackage{booktabs}       % professional-quality tables
\usepackage{amsfonts}       % blackboard math symbols
\usepackage{nicefrac}       % compact symbols for 1/2, etc.
\usepackage{microtype}      % microtypography
\usepackage{amsmath,graphicx,amssymb,xcolor,multirow}
\usepackage{xcolor}

\def\v#1{\mathbf{#1}}
\def\l#1{\mathcal{#1}}

\newcommand{\bowen}[1]{\textcolor{blue}{#1 --Bowen }}
\newcommand{\ming}[1]{\textcolor{red}{#1 --Ming }}

\renewcommand{\bowen}[1]{}
\renewcommand{\ming}[1]{}

\title{Compression of Acoustic Event Detection Models with Low-rank Matrix Factorization and Quantization Training}

\author{
 Bowen Shi \\
 Toyota Technological Institute At Chicago\\
  \texttt{bshi@ttic.edu} \\
 \And
  Ming Sun \\
  Amazon \\
  \texttt{mingsun@amazon.com} \\
  \AND
  Chieh-Chi Kao \\
  Amazon \\
  \texttt{chiehchi@amazon.com} \\
  \And
  Viktor Rozgic \\
  Amazon \\
  \texttt{rozgicv@amazon.com} \\
  \And
  Spyros Matsoukas \\
  Amazon \\
  \texttt{matsouka@amazon.com} \\
  \And
  Chao Wang \\
  Amazon \\
  \texttt{wngcha@amazon.com} \\
}

\begin{document}
% \nipsfinalcopy is no longer used

\maketitle

\vspace{-0.2in}
\begin{abstract}
  %% Deep neural network is a commonly used model architecture in acoustic event detection (AED).
  In this paper, we present a compression approach based on the combination of low-rank matrix factorization and quantization training, to reduce complexity for neural network based acoustic event detection (AED) models. Our experimental results show this combined compression approach is very effective. For a three-layer long short-term memory (LSTM) based AED model, the original model size can be reduced to $1\%$ with negligible loss of accuracy. Our approach enables the feasibility of deploying AED for resource-constraint applications.
\end{abstract}

\vspace{-0.2in}
\section{Introduction}
\label{sec:intro}

Acoustic event detection (AED), the task of detecting whether certain events occur in an audio clip, can be applied in many industry applications [1,2,3,23]. The accuracy of AED models have been increased in a large scale in recent years based on deep learning approaches. However, to ensure high performance, those models are of large scale computation and memory intense, which makes it a challenge to deploy for real industrial applications with limited computation resources and memory. Our paper is focused on increasing the computation efficiency for AED models while maintaining their accuracies, so that AED deployment for resource-constraint industrial applications is feasible.

Compression of neural networks has been explored in broad context. We focus on two widely used and effective methods for deep models: low-rank matrix factorization and and quantization.
%% Low-rank matrix factorization aims at estimating informative parameters by matrix decomposition to reduce weight matrices parameters and multiplications in inference.
Singular value decomposition (SVD) is a common factorization technique and has been explored in feedforward networks [9,10,21,22] and recurrent neural networks (RNN) [11]. Neural network quantization refers to compressing the original network by reducing number of bits required to represent weight matrices, and it has been studied for different model architectures [12,13,14,15,16,19,20]. By reducing the bit-width of weights, model size is reduced, and it also brings considerable acceleration via efficient low bit-width arithmetic operations supports available on hardware. For the quantization approach, It is important to fine-tune models with quantized weights to reduce the performance loss with quantized networks. Here we refer quantization with fine-turning as quantization training.

\section{Methods}
\label{sec:method}

We start by formulating the multi-class audio event detection problem. Given an audio signal $I$ (e.g. log mel-filter bank energies (LFBEs)), the task is to train a model $\v f$ to predict a multi-hot vector $\v y\in\{0, 1\}^C$, with $C$ being the size of event set $\l E$ and $y_c$ being a binary indicator whether event $c$ is present in $I$. We denote $\l D_L=\{(I, \v y)\}$ as the labeled training dataset. Model $\v f$ is trained using cross-entropy loss: $L=-\displaystyle\sum_{(I, \v y)\in \l D_L}\displaystyle\sum_{c=1}^C\{w_c y_c\log f_c(I)+(1-y_c)\log(1-f_c(I))\}$, where $w_c$ is the penalty of positive mis-classification of class $c$. Specifically we focus on the RNN-based model in this paper. Compared to CNNs, it is more memory efficient and easier to deploy on devices with constrained resources.

%% $w_c$ is tuned to balance losses computed from positive and negative instances for each individual event.

%(see eq \ref{eq:sup_loss})
%\begin{equation}
%\centering
%\label{eq:sup_loss}
%L=-\displaystyle\sum_{(I, \v y)\in \l D_L}\displaystyle\sum_{c=1}^C\{w_c y_c\log f_c(I)+(1-y_c)\log(1-f_c(I))\}
%\end{equation}

\textbf{Low-rank matrix factorization} The factorization of weight matrices is based on the SVD compression of LSTM [11].
Let $W_x^l$ and $W_h^l$ denote the input and recurrent matrix of layer $l$, respectively. We first run SVD on $W_h^l$ to retain the top-$r$ singular values and singular vectors, and its top-$r$ right singular vectors $V_h^{lr}$ are shared with input weight matrix (see equation \ref{eq:svd_decomp})
%% The objective is to obtain low-rank factorization of $W_x^l$ and $W_h^l$. 

\vspace{-0.2in}
\begin{equation}
  \label{eq:svd_decomp}
  \begin{split}
    & W_h^l = U_h^l\Sigma_h^l {V_h^l}^T \approx Z_h^l\tilde{V_h^{lr}}^T, \text{where } Z_h^l=\tilde{U_h^{lr}}\tilde{\Sigma_h^{lr}} \\
    & W_x^l \approx Z_x^l\tilde{V_h^{lr}}^T, \text{where }Z_x^l = argmin_{Y} ||Y\tilde{V_h^{lr}}^T-W_x^l || \\
  \end{split}
\end{equation}

\textbf{Quantization training} Quantization refers to representing floating-point values with n-bit integers ($n<32$). as formulated in \ref{eq:quantization}. Note the scaling factor $\alpha$ and minimum value $\beta$ in equation \ref{eq:quantization} are not quantized.
%% A common quantization process including (1). scaling, which scales vectors of arbitrary range to values in $[0, 1]$, (2). quantizing, which rounds any scaled value, and (3). recovering, which scales quantized values back to the original range, 

\vspace{-0.3in}
\begin{equation}
\begin{split}
\label{eq:quantization}
& \hat{\v V} = \frac{\v V-\beta}{\alpha},\ \hat{Q}_{n}(\hat{\v V}) = \frac{[\hat{\v V}(2^n-1)]}{2^n-1},\ Q_n(\v V) = \alpha\hat{Q}_{n}(\hat{\v V})+\beta \\
& \text{with}\ \alpha = \max_{i}{V_i}-\min_i{V_i},\ \beta=\min_i{V_i} \\
\end{split}
\end{equation}

%\begin{equation}
%\label{eq:quantization}
%\begin{split}
%& \hat{\v V} = \frac{\v V-\beta}{\alpha} \\
%& \hat{Q}_{n}(\hat{\v V}) = \frac{[\hat{\v V}(2^n-1)]}{2^n-1} \\
%  & Q_n(\v V) = \alpha\hat{Q}_{n}(\hat{\v V})+\beta \\
%  & \alpha = \max_{i}{V_i}-\min_i{V_i},\ \beta=\min_i{V_i} \\
%\end{split}
%\end{equation}
\vspace{-0.1in}
As the quantization function ($\hat{Q}_n$) is discrete, its gradient is almost zero everywhere. To solve this problem, we apply straight-through estimator [18] to approximate the gradient of full-precision parameter ($\frac{\partial l}{\partial \v V}$) with gradient of quantized value ( $\frac{\partial l}{\partial \hat{\v V}}$) in the fine-tuning. 

To combine low-rank matrix factorization and quantization training, the quantization operator $\hat{Q}_n$ is applied to $Z_h^l$, $Z_x^l$ and $\tilde{V_h^{lr}}$. We quantize both model parameters and inputs. The original RNN is first trained in full-precision until convergence. After the low-rank matrix factorization is applied (equation \ref{eq:svd_decomp}), the model is quantized and fine-tuned with quantization training. We find the fine-tuning step is important to maintain performance with quantization.

\vspace{-0.1in}
\section{Experiments}
The dataset used in our experiments is a subset of Audioset\ [17], which includes a large collection of 10-second audio segments for 632 categories of events. In particular, we select sounds of dog, baby crying and gunshots as our target events. They include both human and non-human vocals, as well as different sound event durations. The three events included in Audioset amount to 13,460, 2,313 and 4,083 audio segments, respectively. We also randomly select 36,036 examples from all other audio clips in Audioset as negative samples. The whole selected dataset is randomly split for training (70$\%$), validation (10$\%$) and test (20$\%$) for each target events. 64 dimensional LFBE features are extracted for each audio clip, with window size of 25 ms and hop size of 10 ms, which are further normalized by global cepstral mean and variance normalization (CMVN).

Our baseline model is a three-layer LSTM with 256 hidden units in each layer. Dropout is added between layers at rate of 0.2. Adam optimizer is used with learning rate fixed at 0.001. The evaluation is based on detection error tradeoff (DET) curve (false negative rate vs. false positive rate). We compute area under curve (AUC) and equal error rate (EER) as the two quantitative measures.

As the distribution of weight matrices' eigenvalues can be different across different LSTM layers, we follow the practice of [11] to set the same threshold $\tau$ across layers as the fraction of retained singular values, defined as $\tau=\displaystyle\sum_{j=1}^r{\sigma_j^l}^2/\displaystyle\sum_{j=1}^N{\sigma_j^l}^2, \sigma_1^l\geq\sigma_2^l\geq...\geq\sigma_{N}^l$.

%\begin{equation}
%  \label{eq:def_tau}
%  $\tau=\frac{\displaystyle\sum_{j=1}^r{\sigma_j^l}^2}{\displaystyle\sum_{j=1}^N{\sigma_j^l}^2}, \sigma_1^l\geq\sigma_2^l\geq...\geq\sigma_{N}^l$
%\end{equation}

Table \ref{tab:svd_fp} summarizes the results of low-rank matrix factorization compared to our baseline 3-layer LSTM. There is no accuracy degradation when $\tau$ is reduced to 0.6, which we hypothesize to be related to the regularizing effects. Table \ref{tab:fr_qt} summarizes the results with quantization compared to our baseline. Post-mortem (PM) refers to the case that  quantization is only applied during inference, while quantization training (QT) refers to the case model fine-tuning is further performed on quantized weights. Our quantization training approach outperforms PM significantly for the 4-bit quantization case. We also note the simple PM quantization preserves the accuracy well (3.0\% drop in AUC and 2.7\% drop in EER) with 8-bit quantization.

\begin{table}[h]
  \setlength{\tabcolsep}{2.5pt}
  \centering
  \caption{\label{tab:svd_fp}AUCs and EERs on test set with different $\tau$ on full-precision low-rank matrix factorization of three-layer LSTM. Lower AUC and EER indicate better performance.}
  \begin{tabular}{cc||cccc|cccc}
    \multicolumn{2}{c||}{Low-rank, Full-precision} & \multicolumn{4}{c|}{AUC (\%)} & \multicolumn{4}{c}{EER (\%)} \\ \hline
    $\tau$ & Params (MB) & Dog & Baby & Gunshot & \bf Avg & Dog & Baby & Gunshot & \bf Avg \\
    1.0 & 5.273 & 7.49 & 6.34 & 5.10 & 6.31 & 15.19 & 13.58 & 11.79 & 13.52 \\
    0.9 & 2.396 & 7.22 & 7.08 & 4.78 & 6.36  & 14.54 & 15.23 & 10.96 & 13.54 \\
    0.8 & 1.069 & 7.61 & 6.36 & 4.88 & 6.28  & 15.26 & 13.52 & 11.05 & 13.27 \\
    0.6 & 0.316 & 7.74 & 6.71 & 5.07 & 6.51 & 15.54 & 13.8 & 11.29 & 13.54 \\
    0.4 & 0.132 & 10.79 & 10.58 & 10.87 & 10.75  & 19.39 & 18.72 & 18.74 & 18.95 \\ \hline
  \end{tabular}
\end{table}

\begin{table}[h]
  \vspace{-0.2in}
  \centering
  \setlength{\tabcolsep}{2pt}
  \caption{\label{tab:fr_qt}AUCs and EERs on test set w/ different quantization bits on three-layer LSTM without low-rank factorization. Lower AUC and EER indicate better performance.}
  \begin{tabular}{cc||cccc|cccc}
    \multicolumn{2}{c||}{Full-rank, Quantization} & \multicolumn{4}{c|}{AUC (\%)} & \multicolumn{4}{c}{EER (\%)} \\ \hline
    \# bits, type & Params (MB) & Dog & Baby & Gunshot & \bf Avg & Dog & Baby & Gunshot & \bf Avg \\
    Full-precision & 5.273 & 7.49 & 6.34 & 5.10 & 6.31 & 15.19 & 13.58 & 11.79 & 13.52 \\
    8-bit PM & 1.318 & 7.73 & 6.52 & 5.18 & 6.48 & 15.43 & 13.64 & 12.39 & 13.89 \\
    8-bit QT & 1.318 & 7.91 & 6.42 & 5.17 & 6.50 & 15.88 & 13.55 & 12.39 & 13.94 \\
    4-bit PM & 0.659 & 10.14 & 9.51 & 12.21 & 10.62 & 18.67 & 17.69 & 21.37 & 19.24 \\
    4-bit QT & 0.659 & 8.15 & 8.08 & 6.05 & 7.43 & 15.91 & 15.45 & 13.08 & 14.81 \\ \hline
  \end{tabular}
\end{table}

Finally, we combine both low-rank matrix factorization and quantization training. Its results are summarized in table \ref{tab:svd_qt}. The attained singular value ratio is fixed to be 0.6, as parameter size and accuracy is well balanced at this point according to table \ref{tab:svd_fp}. When the model is quantized to 8-bit QT ($\approx 1.5\%$ of original size), AUC is only increased by 0.2$\%$ and EER is even improved by a small margin (1.9$\%$). Performance is significantly degraded for 8-bit PM, which is related to the relatively highly compactness and unbounded intermediate outputs in the low-rank setting. Fine-tuning the quantized model is an essential step to reduce this effect. Though the performance is decreased for the 4-bit QT (less than $1\%$ of original model size), its relative degradation is much smaller compared to the 4-bit PM.

\begin{table}[h]
  %% \vspace{-0.2in}
  \centering
  \setlength{\tabcolsep}{2pt}
  \caption{\label{tab:svd_qt}AUCs and EERs on test set with different quantization bits on low-rank factorized three-layer LSTM, $\tau=0.6$. Lower AUC and EER indicate better performance.}
  \begin{tabular}{cc||cccc|cccc}
    \multicolumn{2}{c||}{Low-rank, Quantization} & \multicolumn{4}{c|}{AUC (\%)} & \multicolumn{4}{c}{EER (\%)} \\ \hline
    \# bits, type & Params (MB) & Dog & Baby & Gunshot & \bf Avg & Dog & Baby & Gunshot & \bf Avg \\
    Baseline & 5.273 & 7.49 & 6.34 & 5.10 & 6.31 & 15.19 & 13.58 & 11.79 & 13.52 \\
    8-bit PM & 0.079 & 9.74 & 8.28 & 6.61 & 8.21 & 18.18 & 15.55 & 12.89 & 15.54 \\
    8-bit QT & 0.079 & 7.37 & 6.38 & 5.24 & 6.33 & 15.11 & 13.34 & 11.31 & 13.26 \\
    4-bit PM & 0.040 & 15.26 & 20.89 & 12.98 & 16.38 & 23.55 & 26.38 & 20.53 & 23.48 \\
    4-bit QT & 0.040 & 9.10 & 8.13 & 7.29 & 8.17 & 17.05 & 14.82 & 12.97 & 14.94 \\ \hline
  \end{tabular}
\end{table}

\vspace{-0.2in}
\section{Conclusion}

In this paper we present a simple yet effective compression technique combining low-rank matrix factorization and quantization training. The proposed technique is applied to a multi-layer LSTM for AED. It compresses the AED model size to less than $1\%$ of the original, with AUC and EER performance well maintained. Model fine-tuning with quantization (QT) outperforms a naive quantization scheme (PM), which shows its consistent advantage with different number of quantization bits. 

%% \textcolor{red}{Could we add a few more references for AED and compression, and update arXiv to the actual conf/journal if possible}

%% \textcolor{red}{We need to further reduce content to 3 pages, maybe by reducing margin/font etc?}

\section*{References}

% References follow the acknowledgments. Use unnumbered first-level
% heading for the references. Any choice of citation style is acceptable
% as long as you are consistent. It is permissible to reduce the font
% size to \verb+small+ (9 point) when listing the references. {\bf
%   Remember that you can use more than eight pages as long as the
%   additional pages contain \emph{only} cited references.}
% \medskip

\small

[1] M. Cristani, M. Mecego, and V. Murino\ (2007), Audio-visual event recognition in surveillance video sequences,{\it IEEE Transactions on Multimedia}.

[2] G. Valenzise, L. Gerosa, M. Tagliasacchi, F. Antonacci, and A. Sarti \ (2007) Scream and gunshot detection and localization for audio-surveillance systems,{\it IEEE Conference on Advanced Video and Signal Based Surveillance}.

[3] P. Cano, M. Koppenberger, and N. Wack\ (2005), Content- based music audio recommendation, in {\it Proceedings of the 13th Annual ACM International Conference on Multimedia}.

[4] G. Hinton, L. Deng, D. Yu, G.E. Dahl, A.R. Mohamed, N. Jaitly, A. Senior, V. Vanhoucke, P. Nguyen, T.N. Sainath, and B. Kingsbury\ (2012). Deep neural networks for acoustic modeling in speech recognition: The shared views of four research groups, {\it IEEE Signal processing magazine}.

[5] A. Krizhevsky, I. Sutskever, and G.E.Hinton\ (2012). Imagenet classification with deep convolutional neural networks, in {\it NIPS}.

[6] Y. Aytar, C. Vondrick, and A. Torralba\ (2016). Soundnet: Learning sound representations from unlabeled video, in {\it NIPS}. 

[7] A. Jansen, M. Plakal, R. Pandya, D. Ellis, S. Hershey, J. Liu, R. Moore, and R. Saurous\ (2018). Unsupervised learning of semantic audio representations, in {\it ICASSP}.

[8] Y. Wu, and L. Tan\ (2018). Reducing model complexity for dnn based large-scale audio classification, in {\it ICASSP}.

[9] J. Xue, J. Li and Y. Gong\ (2013). Restructuring of deep neural network acoustic models with singular value decomposition. in {\it Interspeech}

[10] Y. Wang, J. Li, and Y. Gong (2015) Small-footprint high-performance deep neural network-based speech recognition using split-VQ. in {\it ICASSP}

[11] R. Prabhavalkar, O. Alsharif, A. Bruguier, I. McGraw\ (2016). On the compression of recurrent neural networks with an application to lvcsr acoustic modeling for embedded speech recognition. in {\it ICASSP}

[12] I. Hubara, M. Courbariaux, D. Soudry, R. El-Yaniv, and Y. Bengio\ (2018). Quantized neural networks: training neural networks with low precision weights and activations. in {\it Journal of Machine Learning Research, vol. 18, pp. 1-30}.

[13] Q. He, H Wen, S. Zhou, Y. Wu, C. Yao, X. Zhou and Y. Zou\ (2016). Effective quantization methods for recurrent neural networks. {\it arXiv preprint arXiv:1611.10176}

[14] I. Hubara, D. Soudry, and R. El Yaniv\ (2016). Binarized neural networks. {\it arXiv preprint arXiv:1602.02505}

[15] R. Alvarez, R. Prabhavalkar, and A. Bakhtin\ (2016). On the efficient representation and execution of deep acoustic models. in {\it Interspeech}

[16] S. Han, H. Mao, and W. J. Dally\ (2015). Deep compression: Compressing deep neural networks with pruning, trained quantization and huffman coding. {\it arXiv preprint arXiv:1510.00149}

[17] F. Gemmeke, D.P.W. Ellis, D. Freedman, A. Jansen, W. Lawrence, R. C. Moore, M. Plakal, and M. Ritter\ (2017). Audio set: an ontology and human-labeled dataset for audio events, in {\it ICASSP}.

[18] Y. Bengio, N. Leonard and A.C. Courville\ (2013). Estimating or propagating gradients through stochastic neurons for conditional computation. In {\it CoRR, abs/1308.3432}

[19] C. Xu, J. Yao, Z. Lin, W. Ou, Y. Cao, Z. Wang and H. Zha\ (2018). Alternating Multi-bit Quantization for Recurrent Neural Networks, in {\it ICLR}

[20] S. Zhou, Y. Wang, H. Wen, Q. He and Y. Zou\ (2017) Balanced Quantization: An Effective and Efficient Approach to Quantized Neural Networks, in {\it Journal of Computer Science and Technology}

%% [21] G. Tucker, M. Wu, M. Sun, S. Panchapagesan, G. Fu and S. Vitaladevuni\ (2016) Model compression applied to small footprint keyword spotting, in {\it_ICLR}

%% [22] M. Sun, D. Snyder, Y. Gao, V. Nagaraja and M. Rodehorst\ (2017) Compressed time delay neural network for small footprint keyword spotting, in {\it_Interspeech}
[21] G. Tucker, M. Wu, M. Sun, S. Panchapagesan, G. Fu and S. Vitaladevuni\ (2016) Model compression applied to small footprint keyword spotting, in {\textit{Interspeech}}

[22] M. Sun, D. Snyder, Y. Gao, V. Nagaraja and M. Rodehorst\ (2017) Compressed time delay neural network for small footprint keyword spotting, in {\textit{Interspeech}}

[23] C-C. Kao, W. Wang, M. Sun and C. Wang\ (2018) R-CRNN: Region based Convolutional Recurrent Neural Network for Audio Event Detection, in {\it Interspeech}

%% [24] G. Tucker, M. Wu, M. Sun, S. Panchapagesan, G. Fu and S. Vitaladevuni\ (2016) Model compression applied to small footprint keyword spotting, in {\it_Interspeech}

%% [1] Alexander, J.A.\ \& Mozer, M.C.\ (1995) Template-based algorithms
%% for connectionist rule extraction. In G.\ Tesauro, D.S.\ Touretzky and
%% T.K.\ Leen (eds.), {\it Advances in Neural Information Processing
%%   Systems 7}, pp.\ 609--616. Cambridge, MA: MIT Press.

%% [2] Bower, J.M.\ \& Beeman, D.\ (1995) {\it The Book of GENESIS:
%%   Exploring Realistic Neural Models with the GEneral NEural SImulation
%%   System.}  New York: TELOS/Springer--Verlag.

%% [3] Hasselmo, M.E., Schnell, E.\ \& Barkai, E.\ (1995) Dynamics of
%% learning and recall at excitatory recurrent synapses and cholinergic
%% modulation in rat hippocampal region CA3. {\it Journal of
%%   Neuroscience} {\bf 15}(7):5249-5262.
\end{document}